\documentstyle[prb,aps,amsfonts,amssymb,floats,epsfig]{revtex}

\begin{document}
\draft
\twocolumn[\hsize\textwidth\columnwidth\hsize\csname
@twocolumnfalse\endcsname
\title{
Observation of two-magnon bound states in the
two-leg ladders of (Ca,La)$_{14}$Cu$_{24}$O$_{41}$
}
\author{
M. Windt,$^{1}$ M. Gr\"{u}ninger,$^{1}$ T. Nunner,$^{2}$
C. Knetter,$^{3}$ K. Schmidt,$^{3}$ G.S. Uhrig,$^{3}$
T. Kopp,$^{2}$ A. Freimuth,$^{1}$
\\
U. Ammerahl,$^{1,5}$ B. B\"{u}chner,$^{1,4}$ and A. Revcolevschi$^{5}$}
\address{
$^1$\it II. Physikalisches Institut, Universit\"{a}t zu K\"{o}ln,
50937 K\"{o}ln, Germany
\\
{\rm $^2\!$} Experimentalphysik VI, Universit\"{a}t Augsburg,
86135 Augsburg, Germany
\\
{\rm $^3\!$} Institut f\"{u}r Theoretische Physik, Universit\"{a}t zu K\"{o}ln,
50937 K\"{o}ln, Germany
\\
{\rm $^4\!$} II. Physikalisches Institut, RWTH-Aachen, 52056 Aachen, Germany
\\
{\rm $^5\!$} Laboratoire de Chimie des Solides,
Universit\'e Paris-Sud, 91405 Orsay C\'edex, France
}
\date{March 20, 2001}
\maketitle
\begin{abstract}
Phonon-assisted 2-magnon absorption is studied at T=4 K in the spin-1/2 2-leg ladders of
Ca$_{14-x}$La$_{x}$Cu$_{24}$O$_{41}$ ($x$=5 and 4) for $E \! \parallel \! c$ (legs) and
$E \! \parallel \! a$ (rungs). Two peaks at about 2140 and 2800 cm$^{-1}$ reflect van-Hove
singularities in the density of states of the strongly dispersing 2-magnon singlet bound
state, and a broad peak at $\approx $ 4000 cm$^{-1}$ is identified with the 2-magnon continuum.
Two different theoretical approaches (Jordan-Wigner fermions and perturbation theory)
describe the data very well for $J_\parallel \! \approx \! 1050 - 1100$ cm$^{-1}$,
$J_\parallel/J_\perp \! \approx \! 1 - 1.1$.
A striking similarity of the high-energy continuum absorption of the ladders and of
the undoped high $T_c$ cuprates is observed.
\end{abstract}
\vskip2pc]
\narrowtext

Low-dimensional quantum spin systems display a fascinating variety of
low-energy excitations.
Prominent examples are the fractional quantum states of one-dimensional (1D)
chains, the $S$=1/2 spinons, or the variety of bound states in gapped
spin liquids such as, {\em e.g.}, dimerized chains or even-leg ladders.
In 2D, the magnetic excitations of the undoped high-$T_c$ cuprates are usually
discussed in more conventional terms, namely renormalized spin waves.
But these fail to describe the short-wavelength high-energy excitations
probed by optical 2-magnon-plus-phonon absorption.\cite{grueninger}
The nature of incoherent high-energy excitations is currently under
intensive debate.\cite{ho,aeppli,sandviksingh,laughlin}

In this light, the study of other cuprates with related topologies is very instructive.
As far as high-energy excitations are concerned, the Cu$_2$O$_3$ 2-leg ladders realized
in (Ca,La)$_{14}$Cu$_{24}$O$_{41}$ (Ref.\ \onlinecite{structure}) provide a bridge
between 1D physics and the 2D CuO$_2$ layers. Antiferromagnetic $S=1/2$ 2-leg Heisenberg
ladders are represented by the Hamiltonian
\begin{equation}
{\cal H}=\sum_{i} \left\{
J_\parallel ({\bf S}_{1,i}{\bf S}_{1,i+1} + {\bf S}_{2,i}{\bf S}_{2,i+1})
+ J_\perp {\bf S}_{1,i}{\bf S}_{2,i} \right\}
,
\label{hamilton}
\end{equation}
where $J_\perp$ and $J_\parallel$ denote the rung and leg couplings, respectively.
For $J_\parallel$=0 one can excite local rung singlets to rung triplets which become
dispersive on finite $J_\parallel$. For $J_\perp$=0 the $S$=1 chain excitations decay
into asymptotically free $S$=1/2 spinons. An  intuitive picture of  the ``magnons''
(elementary triplets) for $J_\perp$, $J_\parallel \neq$ 0 can be obtained from both
limits: the elementary excitations are either dressed triplet excitations or pairs of
bound spinons with a finite gap $\Delta$ as long as $J_\perp \! > \! 0$.
In a gapped system it is particularly interesting whether bound states occur.
Theoretical studies of 2-leg ladders show that both singlet and  triplet 2-magnon bound
states always exist.\cite{uhrigschulz,damle,sushkovPRL,jurecka,monienPRL,zheng}
However, in the $S$=1/2 copper oxides their experimental observation is a difficult task.
Inelastic neutron scattering directly probes the spin gap, but cannot determine the
high-energy excitations to a sufficient extent due to the large exchange
interactions.\cite{matsudamikeska} Magnetic Raman scattering \cite{sugai} is restricted to
$k$=0 excitations, but for the relevant values of $J_\parallel/J_\perp$ the singlet bound
state appears only for finite wave vector (see Fig.~3 and Ref.~\onlinecite{zheng}).
In this case, optical spectroscopy is the appropriate tool.\cite{jurecka}
Two-magnon-plus-phonon absorption \cite{lorenzana} is able to probe magnetic
excitations throughout the entire Brillouin zone (BZ), since the simultaneous excitation
of a phonon takes care of momentum conservation.

In this paper,
we present optical conductivity data of Ca$_{14-x}$La$_x$Cu$_{24}$O$_{41}$ and identify
the 2-magnon singlet bound state by {\em two} peaks reflecting van-Hove singularities
in the density of states (DOS) of the bound state.
Our theoretical results for bound states in the experimentally
relevant parameter range are based on two different approaches,
namely Jordan-Wigner fermions \cite{jordan} and perturbation expansion about the strong dimer
coupling limit up to 13th order using unitary transformations.\cite{wegner,knetter}
Both yield an excellent description of the optical data for
$J_\parallel \! \approx \! 1050 - 1100$ cm$^{-1}$ and
$J_\parallel/J_\perp \approx 1 - 1.1$.

Single crystals with $x$=5 and 4 were grown by the travelling solvent floating zone
method.\cite{udo}
Their single phase structure and stoichiometry have been verified by x-ray, energy dispersive
x-ray and thermogravimetric analyses.\cite{udo}
The La content $x$ determines the average oxidation state of Cu.
A nominally undoped sample is obtained for $x$=6, which probably is beyond the
La solubility limit.\cite{udo} Single phase crystals could only be synthesized for
$x \leq 5$.
The samples with $x$=5 and 4 on average contain $n$=1/24 and 2/24 holes per Cu,
respectively. But x-ray absorption data show that at least for $n \leq 4/24$ the
holes are located within the chains,\cite{nuecker}
which agrees with previous considerations.\cite{mizuno,carter}
Thus we consider the ladders to be undoped which is supported by the similarity
of our results for $x$=5 and $x$=4.

\begin{figure}[t]
\centerline{\psfig{figure=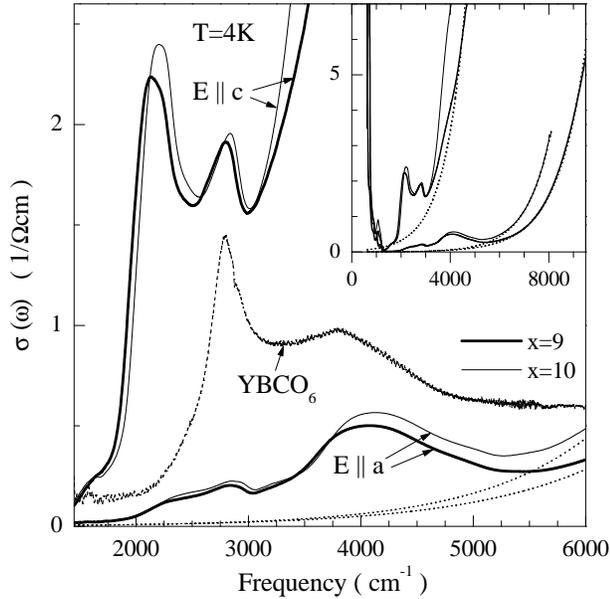,width=8cm,clip=}}
  \caption{Optical conductivity $\sigma(\omega)$ of Ca$_{14-x}$La$_{x}$Cu$_{24}$O$_{41}$
  (thick lines: $x$=5; thin solid lines: $x$=4) for $E\!\parallel\! c$ (legs) and
  $E\!\parallel\! a$ (rungs) at 4~K.\@
  Dashed: Bimagnon-plus-phonon absorption of YBa$_2$Cu$_3$O$_6$ is given for
  comparison.$^1$   
  Inset: (Ca,La)$_{14}$Cu$_{24}$O$_{41}$ data on a larger scale.
  Dotted: estimate of the electronic background (exponential fits for $\omega > 6500$ cm$^{-1}$
  for $E\!\parallel\! a$; $\omega > 4500$ cm$^{-1}$ for $E\!\parallel\! c$).
}
\label{exp}
\end{figure}

The optical conductivity $\sigma (\omega)$ was determined by collecting both
transmission \cite{thickness}
and reflection data between 500 and $12\,000$ cm$^{-1}$ on a Fourier spectrometer.
In Fig.\ \ref{exp} we display $\sigma (\omega)$ of Ca$_{14-x}$La$_x$Cu$_{24}$O$_{41}$ at T=4 K
for polarization of the electrical field parallel to the legs, $E \! \parallel \! c$, and to the
rungs, $E \! \parallel \! a$. Phonon and multi-phonon absorption dominates $\sigma(\omega)$ below
$\approx $ 1300 cm$^{-1}$ (see inset). A steep increase of the electronic background, probably
due to
interband excitations of charge-transfer type, is observed above 6000 cm$^{-1}$ for
$E \! \parallel \! a$ (3000 cm$^{-1}$ for $E \! \parallel \! c$). Our analysis focuses on the two
peaks between 2000 and 3000 cm$^{-1}$ and the broad feature at 4000 cm$^{-1}$.
Also plotted in Fig.\ 1 is $\sigma(\omega)$ of YBa$_2$Cu$_3$O$_6$ (Ref.\ \onlinecite{grueninger}),
a typical example of the 2-magnon-plus-phonon absorption spectrum of the undoped 2D
cuprates.\cite{lorenzana}
This comparison gives a first motivation to interpret
the peaks in $\sigma(\omega)$ of the ladders as magnetic excitations. Note that in the ladders
both the exchange constants and the relevant Cu-O bond stretching phonon frequencies are comparable
to those found in the 2D cuprates. Since the exchange coupling in the chains is much smaller,
$J_{\rm chain}\!\approx\! -14$ cm$^{-1}$ (Ref.\ \onlinecite{carter}),
their magnetic excitations do not contribute to $\sigma(\omega)$ in this frequency range.
Reducing the La content $x$ from 5 to 4 causes a slight {\em blue-}shift of the magnetic absorption
that is opposite to the {\em red-}shift of the electronic background.
We attribute the blue-shift to an increase of the exchange constants caused by the reduction of the
lattice parameters (Ca is smaller than La), and the red-shift of the background to an increased hole
density in the {\em chains}.

Since spin is conserved, infrared absorption is sensitive to $\Delta S$=0,  singlet excitations,
{\em e.g.}, the excitation of two $S$=1 magnons coupled to $S_{\rm tot}$=0.
Direct excitation of two magnons is Raman active \cite{sugai} but infrared forbidden due to the
inversion symmetry on a 180$^\circ$ Cu-O-Cu bond. We can effectively avoid this selection rule
by simultaneously exciting a Cu-O bond stretching phonon that breaks the symmetry.\cite{lorenzana}
Hence the lowest order infrared-active magnetic absorption is a 2-magnon-plus-phonon process.
The low values of $\sigma (\omega) \lesssim 2\; \Omega^{-1}$cm$^{-1}$ indeed indicate a weak
higher-order absorption process.

\begin{figure}[t]
\centerline{\psfig{figure=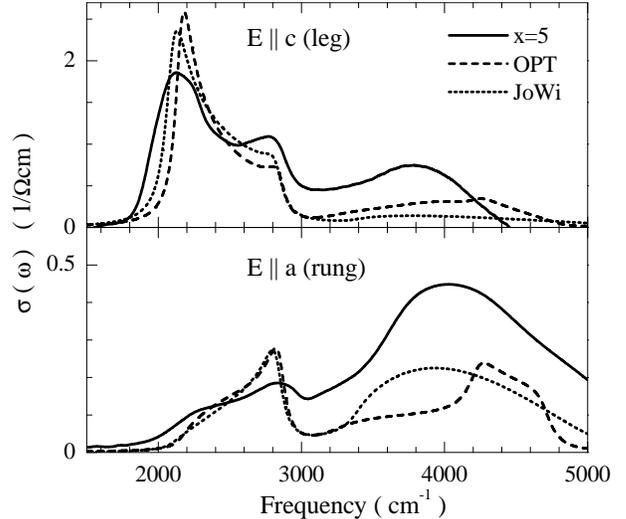,width=8cm,clip=}}
  \caption{Comparison of experiment (solid lines for $x$=5) and theory for
  $J_\perp$ = $J_\parallel$ = 1050 cm$^{-1}$ (OPT: optimized perturbation) and
  1100 cm$^{-1}$ (JoWi: Jordan-Wigner fermions), with $\omega_{ph}$ = 600 cm$^{-1}$.
  The exponential fits of the electronic background have been subtracted from the
  experimental data (dotted lines in Fig.\ \ref{exp}).
}
\label{theo}
\end{figure}

\begin{figure}[t]
\centerline{\psfig{figure=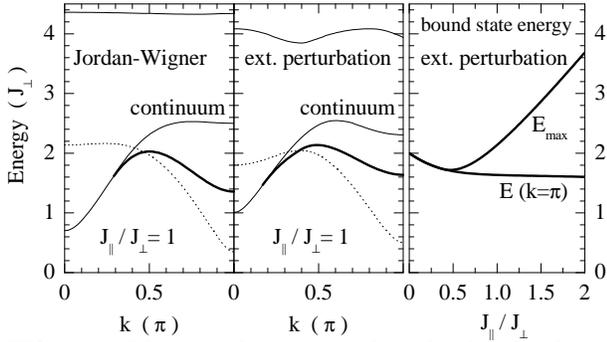,width=8cm,clip=}}
  \caption{1-Magnon dispersion (dotted), band edges of the 2-magnon continuum
  (solid), and bound singlet dispersion (thick lines) for $J_\parallel/J_\perp$=1
  obtained from Jordan-Wigner fermions (left) and extrapolated perturbation (mid
  panel).
  Right: energy of the van-Hove singularities of the bound singlet at
  $k=\pi$ and at the dispersion maximum as a function of $J_\parallel/J_\perp$.
  }
\label{disp}
\end{figure}

The 2-magnon-plus-phonon contribution to $\sigma(\omega)$ was evaluated by Jurecka and
Brenig \cite{jurecka} in the extreme case of strong rung coupling $J_\parallel \rightarrow 0$.
They found a single sharp peak reflecting the singlet bound state.
In order to gain quantitative control over the experimentally relevant coupling regime of
$J_\parallel/J_\perp \approx 1$ we use two different approaches to calculate the dynamical 4-point
spin correlation function, namely Jordan-Wigner fermions and perturbation theory.\cite{control}
In the former, we make use of the Jordan-Wigner transformation \cite{jordan} to rewrite the spins
as fermions with a long-ranged phase factor. Expanding the phase factor yields new interaction terms
between the fermions which we treat in RPA (random phase approximation).
The second approach is perturbative in nature. It is performed by a continuous unitary
transformation.\cite{wegner,knetter} The 1-particle energies and the 2-particle bound state energies
are extrapolated by standard techniques (Pad\'e and Dlog-Pad\'e); the spectral densities are computed
by optimized perturbation.\cite{kleinert}
Details will be given elsewhere.
The magnon (elementary triplet) dispersion, the 2-particle continuum and the dispersion of the
singlet bound state obtained for $J_\perp=J_\parallel$ are given in the two left panels of
Fig.\ \ref{disp}. The results of the two techniques agree well even though there is a quantitative
discrepancy ($\lessapprox 10\%$) in some energies (dispersion at $k$=0 and $\pi$, bound state energy).
The magnon dispersion of extrapolated perturbation (mid panel)
reproduces previous results very well.\cite{johnstonladder,wzheng98}
We focus on the singlet bound state which shows a considerable dip at $k=\pi$.
This has gone unnoted thus far because it occurs only for $J_\parallel/J_\perp \gtrapprox 0.5$
(see right panel in Fig.~3 and Ref.~\onlinecite{monienPRL}).

To compare with experiment we consider the effect of the phonons.
The total spectral weight is obtained by taking into account a dependence
of the exchange constants on the external electric field {\bf E} and the
displacements of O ions ${\bf u}$,
$J_{\parallel,\perp} \equiv J_{\parallel,\perp}({\bf E},{\bf u})$
(Refs.\ \onlinecite{lorenzana,grueninger}).
The phonons modulate the intersite hopping and the on-site energies
on both Cu and O sites. We expand $J({\bf E},{\bf u})$ to order
$d^2\!J/d{\bf u}d{\bf E}$ which
entails the coupling of a photon to a phonon and two neighboring spins.
This determines how to integrate the spin response in the BZ. Here, the weight factor
is a mixture of $\sin^4(k/2)$ (Ref.\ \onlinecite{lorenzanaeder}) and of an isotropic
($k$-independent) form factor, where the former dominates.\cite{weightfactor}
For simplicity we consider a pure form factor $\sin^4(k/2)$ and
Einstein phonons with $\omega_{ph} = 600$ cm$^{-1}$ as is common for the cuprates.
Our findings depend only very weakly on the precise phonon energy.

In Fig.\ \ref{theo} we compare the theoretical results with the experimental data.
For the former an artificial broadening of $J_\perp/20$ is used.
For the latter, the magnetic part of $\sigma(\omega)$ was obtained by subtracting
the electronic background (dotted lines in Fig.\ \ref{exp}).
For $E\!\parallel \! a$, the background was determined unambiguously by an exponential
fit for $\omega \! > \! 6500$ cm$^{-1}$. The $\sigma_a (\omega)$ curves for $x$=5
and 4 nearly coincide after subtraction of the background.
This corroborates the assumption that the ladders are undoped.
For $E\!\parallel\! c$, the measurable range of transmission was limited to below
$\approx 4500$ cm$^{-1}$ by the higher absorption, and the precise shape and spectral
weight of the highest peak at $\approx$ 4000 cm$^{-1}$ cannot be determined
unambiguously.

The two peaks at 2140 and 2800 cm$^{-1}$ ($x$=5) can be identified with the 1D
van-Hove singularities in the DOS of the singlet bound state. The lower peak corresponds
to the singlet energy at the BZ boundary, $E(k=\pi)$, the higher one to the maximum of
the singlet dispersion $E_{\rm max}$ at about $k \approx \pi/2$ (see Fig.\ \ref{disp}).
These two energies determine the two free magnetic parameters $J_\perp$ and $J_\parallel$.
The experimental $\sigma(\omega)$ is well described by both theories
for $J_\parallel \! \approx \!$ 1050 - 1100 cm$^{-1}$ and $J_\parallel/J_\perp \! \approx \! 1$
(see Fig.~2) providing an unambiguous identification of the experimental features.
The quantitative analysis can be pushed a step further \cite{energies} based on the values
of $E_{\rm max}$ and $E(k=\pi)$ computed by extrapolated perturbation (right panel of Fig.\ 3).
The strong dependence of $E_{\rm max}$ on $J_\parallel / J_\perp$ allows to pinpoint
the exchange constants, yielding the same value for $J_\parallel \! \approx \!$ 1050 cm$^{-1}$
and a slightly larger ratio of $J_\parallel/J_\perp \approx 1.1$.
Our interpretation of the experimental features is confirmed by the good agreement between
theory and experiment concerning the line shape of the bound states in $\sigma(\omega)$.
Excellent justification for this interpretation is also provided by the selection rule
stemming from reflection symmetry about the $a$ axis (RS$a$). Both theories show that the
bound singlet at $k=\pi$ is {\it even} under RS$a$. But the excitations at $k=\pi$
are {\it odd} under RS$a$ for $E \!\parallel\! a$ and {\it even} for $E \!\parallel\! c$.
Thus the weight of the bound state varies as $(k - \pi)^2$ for $E \!\parallel\! a$
whereas it is prevailing  for $E\!\parallel\! c$. This explains the low spectral weight
of the lower peak for $E\!\parallel\! a$. It is reduced to a weak shoulder.

Shortcomings of the theory are the overestimation of the spectral weight of
the 2800 cm$^{-1}$ peak for $E\!\parallel\! a$ and that the onset of $\sigma(\omega)$
around 2000 cm$^{-1}$ is sharper than observed experimentally. However, the agreement
is better than one may have expected since we neglected both the frustrating
coupling between neighboring ladders and the ring exchange. A finite inter-ladder
coupling will produce a dispersion of the bound state along $k_a$ and thereby broaden
the features in $\sigma(\omega)$, which can explain a smearing out of the onset at
2000 cm$^{-1}$.
Without the ring exchange, the analysis of experimental data of various other techniques
suggested a larger leg coupling $J_\parallel/J_\perp \gtrapprox 1.5$ (for a detailed
discussion, see Ref.\ \onlinecite{johnstonladder}). One finds a 1-magnon gap of
$\Delta \approx $ 280 cm$^{-1}$ and a dispersion extending up to $\approx $ 1550 cm$^{-1}$
(Ref.\ \onlinecite{matsudamikeska}). For $J_\parallel/J_\perp$=1 the dispersion extends
from $\Delta \approx 0.5 J_\perp$ up to $\approx 2 J_\parallel$ (see Fig.\ \ref{disp}).
One obvious way to reduce $\Delta $ with respect to the maximum is to increase the ratio
$J_\parallel/J_\perp$, in this case to $\gtrapprox$ 1.5. Such a large value is difficult to
reconcile with the microscopic parameters, in particular with the similar Cu-O bond lengths
along the legs and the rung which has provoked a controversial
discussion.\cite{johnstonladder}
Our analysis shows that any values of $J_\parallel/J_\perp$ larger than $1.2$ can be excluded.
Recently, it was pointed out that the neutron data are also consistent with an isotropic exchange
$J_\parallel/J_\perp \approx 1 - 1.1$ and $J_\parallel \approx$ 900 cm$^{-1}$, if a ring exchange
of $\approx $ 0.15 $J_\parallel$ is taken into account.\cite{matsudamikeska}
Considering the fact that the ring exchange will renormalize $J_\parallel$ and $J_\perp$,
this is in perfect agreement with our findings.

Judging the weight of the high-energy continuum one should note that it increases
by taking into account an admixture of an isotropic form factor to the $\sin^4(k/2)$ form
factor used here.
Second, the 2-particle excitations considered here carry only $\approx$ 75\% of the total
spectral weight.
In the 2D cuprates, Lorenzana {\em et al.} \cite{lorenzana99} argued that ring exchange
increases the spectral weight at high energies.
It will be most interesting to check the influence of ring exchange on the continuum of the
ladders, since these provide a technically much better controlled ground.
Note that the theoretical curves in Fig.\ \ref{theo} describe the frequency range of continuum
absorption very well, which gives strong support to our interpretation.

Finally, we address the similarity of $\sigma(\omega)$ of the ladder
(Ca,La)$_{14}$Cu$_{24}$O$_{41}$ and of the undoped 2D cuprates (see Fig.\ \ref{exp}).
The 2D case does not show a truly bound state, the sharp peak at 2800 cm$^{-1}$ in $\sigma(\omega)$
of YBa$_2$Cu$_3$O$_6$ is caused by a resonance, an {\em almost} bound state lying within the
continuum.\cite{lorenzana} This main peak is well described in terms of 2-magnon-plus-phonon
absorption, but the high-energy peak at 3800 cm$^{-1}$ is absent in spin-wave
theory.\cite{grueninger} Note that this discrepancy is particular for the 2D $S$=1/2 case,
the high-energy excitations are absent in the comparable $S$=1 system La$_2$NiO$_4$
(Refs.\ \onlinecite{lorenzana,perkinsNi}).
In the cuprates, the magnetic origin of both peaks has been confirmed by absorption
and Raman measurements under high pressure.\cite{graybeal}
It has been suggested \cite{grueninger} that the high-energy weight in 2D $S$=1/2 compounds
is due to strong quantum fluctuations that go {\em beyond} spin-wave theory.
The intriguing similarity of the 2-particle continuum of the $S$=1/2 quasi-1D ladder
(Ca,La)$_{14}$Cu$_{24}$O$_{41}$ with the 3800 cm$^{-1}$ peak of YBa$_2$Cu$_3$O$_6$
strongly indicates that the high-energy spectral weight is indeed a signature of strong
quantum fluctuations in this compound.

In conclusion, the existence of the singlet bound state and the 2-particle continuum
was demonstrated experimentally in the optical conductivity spectrum $\sigma(\omega)$
of the $S$=1/2 2-leg ladder compound (Ca,La)$_{14}$Cu$_{24}$O$_{41}$.
By two independent theoretical approaches we confirmed that the two sharp peaks plus
the important broad continuum seen in experiment are the unambiguous signature of a
1D dispersive bound singlet with strong incoherent quantum fluctuations.
The experimental spectral weight distribution reflects the theoretical selection rules
perfectly. Quantitative analysis yields $J_\parallel \approx 1050 - 1100$ cm$^{-1}$
and $J_\parallel/J_\perp \approx 1 - 1.1$.
Our findings indicate that the similar experimental results for the high-energy
excitations of the undoped high-$T_c$ materials are also due to strong quantum
fluctuations.

We acknowledge fruitful discussions with E.~M\"{u}ller-Hartmann.
This project is supported by the DFG
in FR 754/2-1, SP 1073 and  SFB 484, by the  BMBF 13N6918/1
and by the DAAD in the frame of PROCOPE.


\begin{references}
\bibitem{thickness}The transmission was measured on samples with a thickness of
44 $\mu$m ($x$=5) and 60 $\mu$m ($x$=4).
\bibitem{control}
Both methods are controlled in the sense that
they become exact on $J_\parallel/J_\perp \to 0$.
\bibitem{weightfactor}
The precise mixture depends on the 3D phonons
and is beyond the physics of an isolated ladder.
However, it does not affect the peak frequencies.
\bibitem{energies}Energies can be extrapolated much more accurately than
spectral densities.
\end{references}
\end{document}